\documentclass[journal=jpclcd,manuscript=article]{achemso}
\usepackage[version=3]{mhchem} % Formula subscripts using \ce{}
\usepackage[T1]{fontenc}       % Use modern font encodings

\usepackage{array}
\usepackage{booktabs}
\usepackage{listings}
\usepackage{lmodern}
\usepackage{microtype}
\usepackage{caption}
\usepackage{float}
\usepackage{natbib}
\author{Kiran Sankar Maiti}
\affiliation{Lehrstuhl f{\"u}r Physikalische Chemie, Technische Universit{\"a}t  M{\"u}nchen, D-85747 Garching, Germany}
\altaffiliation{Current address: Universit\"at Bayreuth, Universit\"atsstra{\ss}e 30, D-95447 Bayreuth, Germany}
\email{kiran.maiti@ch.tum.de}
\phone{+49 (0)921 16835173}
\fax{+49 (0)89 289 13622}
\author{Susmita Roy}
\affiliation{Lehrstuhl f{\"u}r Theoretische Physik, Universit\"at Bayreuth, Universit\"atsstra{\ss}e 30, D-95447 Bayreuth, Germany.}
\author{Christoph Scheurer}
\affiliation{Lehrstuhl f{\"u}r Physikalische Chemie, Technische Universit{\"a}t  M{\"u}nchen, D-85747 Garching, Germany}

\title[An \textsf{achemso} demo]
  {Unexpected red shift of C-H vibrational band of Methyl benzoate}

\abbreviations{CARS, 2DIR, PES}

\begin{document}

%\begin{tocentry}

% \includegraphics[width=0.95\linewidth,angle=0.0]{figTOCjpcl}%\includegraphics[height=3cm]{example}

%Some journals require a graphical entry for the Table of Contents. This should be laid out ``print ready'' so that the sizing of the text is correct.

%Inside the \texttt{tocentry} environment, the font used is Helvetica 8\,pt, as required by \emph{Journal of the American Chemical Society}.

%The surrounding frame is 9\,cm by 3.5\,cm, which is the maximum permitted for  \emph{Journal of the American Chemical Society} graphical table of content entries. The box will not resize if the content is too big: instead it will overflow the edge of the box.

%This box and the associated title will always be printed on a separate page at the end of the document.

%\end{tocentry}

\begin{abstract}
The C-H vibrational bands become more and more important in the structural determination of biological molecules with the development of CARS microscopy and 2DIR spectroscopy. Due to the congested pattern, near degeneracy, and strong anharmonicity of the C-H stretch vibrations, assignment of the C-H vibrational bands are often misleading. Anharmonic vibrational spectra calculation with multidimensional potential energy surface interprets the C-H vibrational spectra more accurately. In this article we have presented the importance of multidimensional potential energy surface in anharmonic vibrational spectra calculation and discuss the unexpected red shift of C-H vibrational band of Methyl benzoate.
\end{abstract}


Vibrational spectroscopy is a valuable experimental tool to understand the structure and dynamics of molecules\cite{faye2013}. Vibrational spectra of biological molecules are often congested and difficult to assign\cite{peuk2016:2312}. In practice, vibrational spectral analysis of biomolecules are dominated by vibrational bands, which can be described approximately in terms of oscillators localized in each respective unit and their mutual coupling. The most extensively studied bands are amide-I and amide-II in the range of 1500 to 1700\,cm$^{-1}$, and amide-A and amide-B between 3000 to 3500\,cm$^{-1}$, for the vibrational spectral analysis of biomolecules\cite{hoch2007:14190}. Recently developed coherent anti Stokes Raman scattering (CARS) microscopy uses the inherent molecular vibration as a probe to image the biological systems, open up a new direction to understand the biological activities even in live cell\cite{zumb1999:4142}. The C-H vibrational bands draw the most interest of CARS microscopist to identify fatty acids in cellular environment\cite{potm2004:40,chen2015:aaa8870}. Not only microscopy, development of two dimensional infrared (2DIR) spectroscopy\cite{borm2000:41}, gives an enormous opportunity to disentangle the congested vibrational bands and allows to study the C-H vibrational bands and their coupling in detail\cite{mait2015:24998}. 
As a results, C-H vibrational bands become an important probe for structural determination of the biological molecules. However, both the experimental methodologies are complicated and require more refined theoretical and computational model to understand the C-H vibrational spectra. \\

The C-H stretching modes are highly localized and  in general it is considered as nearly pure hydrogen motion, essentially insensitive to the molecular conformation and environment\cite{snyd1978:395}. In reality C-H vibrational modes not only interact with themselves but also to the other vibrational bands in the molecule, as a results a strong anharmonic effect is observed. Due to the anharmonic effect, the C-H vibrational spectra calculation within the harmonic approximation is not sufficient. \\

Several attempts have been taken to assign the anharmonic C-H vibrational bands. However, due to the congested vibrational pattern and near degeneracy of the vibrational bands, calculation of the C-H vibrational bands are often troublesome\cite{faim1976:387,yu2007:8971}. The main difficulties of the anharmonic spectra  calculation is that different vibrational bands are not mutually separable. Therefore one has to face the task of calculating wavefunctions and energy levels for systems of many couple degrees of freedom which increases very fast with the system size. To overcome this problem, several attempts have been made, e.g.,  diffusion quantum Monte Carlo (DQMC)\cite{ande1975:1499}, vibrational configuration interaction (CI)\cite{baro2009:540}, vibrational self-consistent field (VSCF)\cite{jung1996:10332}, etc. methods. Among them the VSCF method is the most successful to effectively handle reasonably large molecular systems.\\

The success of the VSCF method  depends upon the accuracy of the calculated potential energy surface (PES)\cite{boun2008:194,roy2013:9468}. In the VSCF approach, the PES enters into the calculation in the form of multidimensional integrals for the effective one-dimensional potentials which is the most difficult part of the VSCF calculations. To simplify these integrals it is possible to express the PES in terms of a hierarchical many-body expansion\cite{cart1997:1179}. 
\begin{eqnarray}V(q_1, \cdots, q_N)&=&\sum_j^N V_j^{(1)}(q_j)+ \sum_{i<j}
  V_{i,j}^{(2)}(q_i,q_j)\nonumber \\
  &+&\sum_{i<j<k}V_{i,j,k}^{(3)}(q_i,q_j,q_k)+\cdots  \nonumber\\ 
&+&\sum_{i<j\cdots<r<s}V_{i,j,\cdots,r,s}^{(n)}(q_i,q_j,\cdots,q_r,q_s)+\cdots.
\label{eq:vscf-13}\end{eqnarray}
where $V_j^{(1)}(q_j)$ is the diagonal potential, $V_{i,j}^{(2)}(q_i,q_j)$ is
the pairwise potential, $V_{i,j,k}^{(3)}(q_i,q_j,q_k)$ is the triple coupling
and so on. 
In typical molecular problems quadruple and higher interaction potentials have a negligible influence on the vibrational spectrum, therefore they are commonly neglected in the VSCF calculations. In a recent detailed analysis it has been shown that even only few triple couplings have a significant influence\cite{rauh2004:9313}. For typically large molecular systems diagonal and pair potentials are usually sufficient if the PESs are calculated accurately in an appropriately chosen coordinate system\cite{mait2013:1100,mait2015:19735}. Additionally diagonal potentials may be calculated employing a higher level computational method than the pair potentials to increase the accuracy of the VSCF calculations significantly. Such a dual level computation not only improves the accuracy but also reduces the computational expenses significantly due to the exclusion of large number of possible pair potential.\\

\begin{figure}[h]
\centering
  \includegraphics[width=0.95\linewidth,angle=0.0]{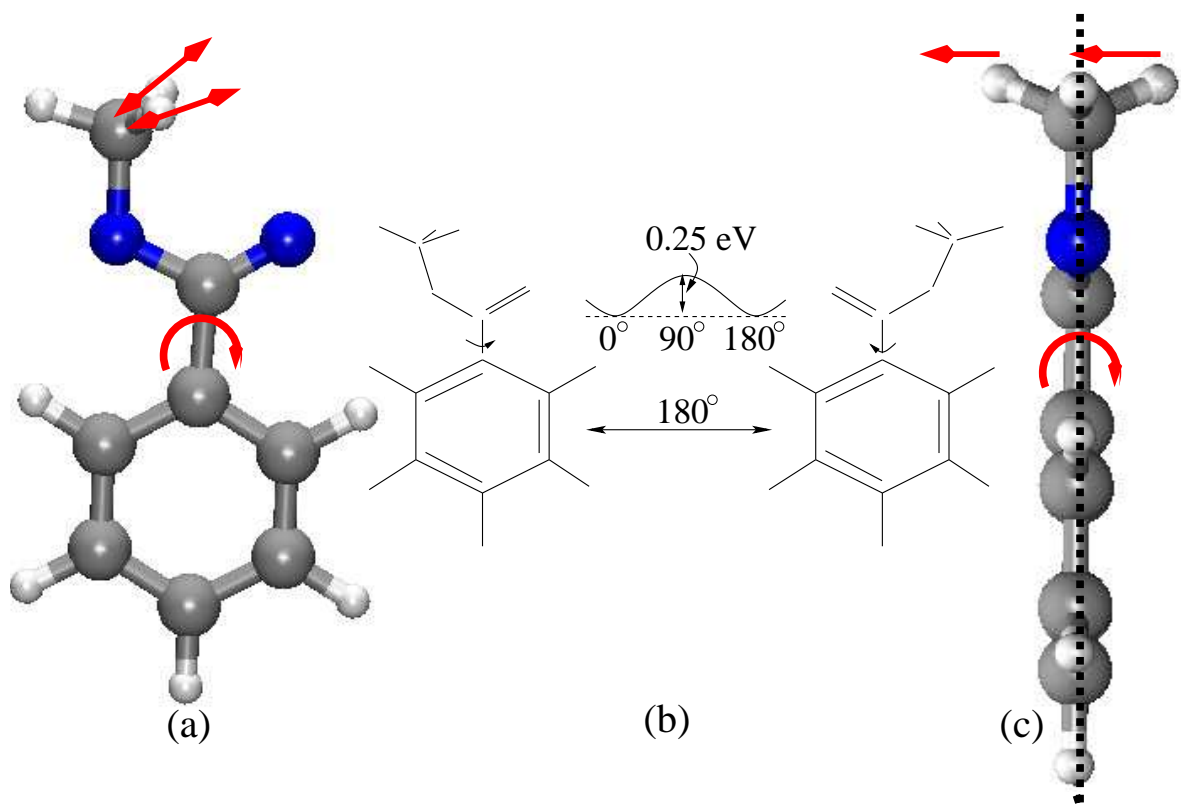}
  \caption{(a) Structure of Methyl benzoate, where plane of the molecule on the paper, (b) The structural conformers of Methyl benzoate and their approximate interconversion barrier height and (c) Structure of Methyl benzoate, where molecular plane is perpendicular to the paper.}
  \label{fig:mb-structute}
\end{figure}

To understand the vibrational coupling between the C-H vibrational bands and other vibrational motions of the molecule, Methyl benzoate has been chosen as a model molecule. In the potential energy minima, the ester group remains in the same plane as the benzene ring. As a result, Methyl benzoate is a planar molecule except for the two hydrogen atoms of the methyl group, which are symmetrically out of plane with respect to the rest of the molecule (Fig.\,\ref{fig:mb-structute}). Due to the very low potential barrier ($\sim$0.25\,eV), ester group can easily rotate around the C-C bond axis. In the present work we have investigated the coupling between this rotational motion of ester group and the C-H vibrations. To keep the C-H vibrational region less congested, the $\beta$-hydrogen has been replaced by deuterium, which shifts one of the C-H vibrational band at around 2200\,cm$^{-1}$.\\

\begin{figure}[h]
\centering
  \includegraphics[width=1.0\linewidth,angle=0.0]{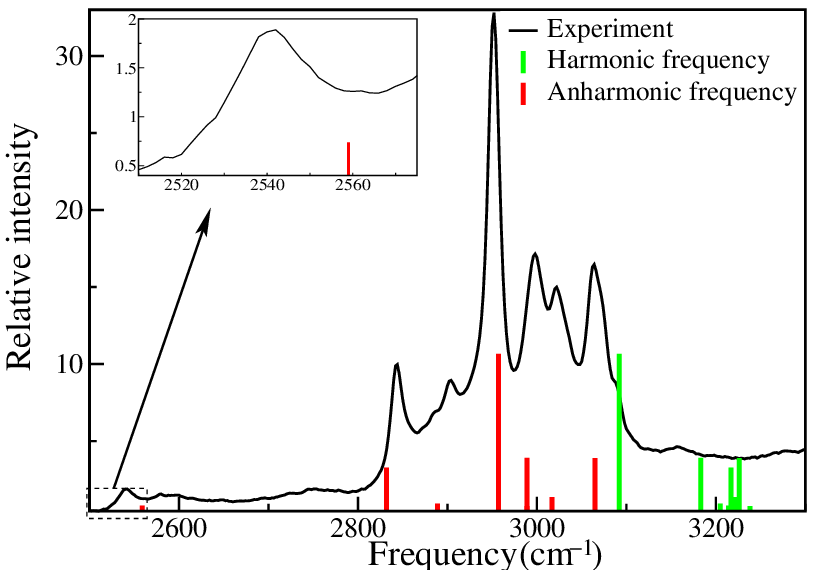}
  \caption{One dimensional FTIR spectra (black line) of deuterated Methyl benzoate plotted along with calculated harmonic (green sticks) and anharmonic (red sticks) vibrational frequencies. The intensity of the calculated harmonic and anharmonic frequencies are obtained from harmonic frequency calculation. The spectrum of the C-H vibrational frequency shift is shown in inset.}
  \label{fig:ch-spec}
\end{figure}

\begin{table}
\caption{The C-H vibrational frequencies and their assignment of deuterated Methyl benzoate with different level of theories, where diagonal grids are calculated with  the DF-MP2/AVTZ level. Mode numbers are based on normal mode frequencies.}\label{tab:final-allH} 
\begin{tabular*}{1.0\textwidth}{@{\extracolsep{\fill}}cccccl}
\hline
Mode&Harmonic & Diagonal& Experiment & Pair coupling & \multicolumn{1}{c}{Assignment\textsuperscript{\emph{a}}}\\
& cm$^{-1}$ & cm$^{-1}$ & cm$^{-1}$ & cm$^{-1}$ & \\
\hline
%40 & 1766  & 1759 & 1915 & 1724 & 1724 & 1778 & 1724 & 1724 & 1724  & C=O s                       \\
{\bf 41} & 3092 & 3059 & 2998 & 2989  & methyl C-H ss\\
{\bf 42} & 3183 & 3255 & 2543 & 2559  & methyl C-H as\\
{\bf 43} & 3205 & 3208 & 2905 & 2889  & phenyl C-H as\\
{\bf 44} & 3214 & 3247 & 2185 & 2171  & phenyl C-D as\\
{\bf 45} & 3217 & 3213 & 2845 & 2832  & methyl C-H ss\\
{\bf 46} & 3222 & 3233 & 2954 & 2957  & phenyl C-H as\\
{\bf 47} & 3226 & 3191 & 3064 & 3065  & phenyl C-H as\\
{\bf 48} & 3238 & 3161 & 3022 & 3017  & phenyl C-H as\\
\hline
\end{tabular*}\\
  \textsuperscript{\emph{a}}Symbols: ss = symmetric stretch, as = asymmetric stretch, methyl C-H = Methyl group C-H band, phenyl C-H = phenyl ring C-H band. 
\end{table}

Anharmonic vibrational frequencies of the C-H vibrational band of Methyl benzoate has been plotted along with harmonic frequencies and FTIR spectra in Fig.\,\ref{fig:ch-spec}. In general the C-H vibrational bands are localized in the region of 2800 - 3100\,cm$^{-1}$ for biological molecules\cite{mait2012:16294}. Calculated harmonic frequencies are localized at 3050\,-\,3250\,cm$^{-1}$, which are quite higher than the experimental frequencies of the C-H vibrations\cite{chat1968:335}. Diagonal anharmonic frequencies, calculated from the anharmonic diagonal PES, are also very high compare to the experimental C-H vibrational frequencies (Table\,\ref{tab:final-allH}). However, the anharmonic vibrational frequencies calculated with the pair coupling potentials are in very good agreement with the experimental spectra and one but all other are localized in the range of 2800 - 3100\,cm$^{-1}$. One of the anharmonic C-H vibrational frequency calculated from the pair coupling potential is unexpectedly red shifted about 400\,cm$^{-1}$. Surprisingly, a small spectral feature has been observed at the experimental spectra, approximately at the same place of the shifted C-H anharmonic frequency. In practice such a small spectral feature does not create any attention and overlooked as an artifact or noise. However, when different level of anharmonic frequency calculations from pair coupling potentials yield the same results, it draws a special interest for further investigation. \\

To verify the existence of this small spectral feature we have calculated the anharmonic vibrational frequencies with different level of quantum chemistry methods and employing different size of basis sets. A very interesting finding  has been observed. All the anharmonic frequency calculations with the pair coupling potentials yield the similar frequency shift ($\sim$\,400\,cm$^{-1}$) for one of the C-H vibrational band. There is no indication of such a large frequency shift neither with the harmonic nor with the anharmonic frequency calculations using diagonal potential. The frequency shift is observed only when pair coupling potential has been used. Therefore there must be one or more vibrational bands which are coupled with one of the C-H vibrational band. A close inspection indicates that the C-H vibrational frequency at 2560\,cm$^{-1}$ originates from the asymmetric C-H stretch vibration in the methyl group. Additionally, it is coupled with the out of plane rotational motion of the ester group with respect to the phenyl ring around the C-C bond.\\

Due to the low potential barrier ($\sim$0.25\,eV) in normal thermal condition ester group rotates around the C-C bond with respect to the phenyl ring. Such a out of plane rotation is pulled one of the out of plane hydrogen atom of methyl group to the plane of the molecule and other hydrogen is pushed from the molecular plane. This bi-directional forces (Fig.\,\ref{fig:mb-structute}a and \ref{fig:mb-structute}b) change the force constant of the anti-symmetric C-H stretch vibration significantly, as a result an unexpected red shift is observed. Therefore it is necessary to account the rotational motion of the ester group for accurate description of the anti-symmetric C-H vibrational band. Since the harmonic and diagonal anharmonic calculations do not account the coupling, it fails to calculate the anti-symmetric vibrational band accurately. The pair coupling potential considers the rotation of ester group and anti-symmetric motion of the methyl C-H vibration together, resulting the more accurate C-H stretch vibration. \\

In this article we have presented the importance of the multi dimensional potential energy surface in anharmonic vibrational spectra calculation. The C-H vibrational bands have been calculated with the harmonic as well as anharmonic approximation. The calculated harmonic and anharmonic diagonal frequencies are significantly higher than the experimental result. The anharmonic C-H vibrational frequencies calculated from the pair coupling potential have very good agreement with the experimental result. This findings indicate that one dimensional potential is not sufficient for the anharmonic vibrational frequency calculation, at least pair coupling potential is necessary for better description of the anharmonic vibrational spectra.\\

{\bf Computational methods:} All computations have been performed using the MOLPRO quantum chemistry program\cite{MOLPRO_brief}. Second order M{\o}ller-Plesset (MP2) perturbation theory\cite{moll1934:618} has been used for the geometry optimization employing augmented correlation-consistence polarized-valence-triple zeta (aug-cc-pVTZ) basis set. Harmonic normal-mode analysis has been performed with the density fitting MP2 (DF-MP2) method using aug-cc-pVTZ regular basis and cc-pVTZ fitting basis set. Diagonal potentials have been calculated with MP2 method using aug-cc-pVTZ basis set. The anharmonic pair coupling potentials are calculated with DF-MP2 method and employing the same cc-pVDZ basis set for regular as well as fitting basis sets. \\

\vspace{-0.5cm}
\begin{acknowledgement}

The authors thank to Deutsche Forschungsgemeinschaft for financial support and to Leibniz-Rechenzentrum for computational facility.

\end{acknowledgement}

%\begin{suppinfo}

%\end{suppinfo}

\bibliography{shiftCHred}

\end{document}